\documentclass[11pt]{article}
\usepackage{moriond,epsfig}

\def\be{\begin{equation}}
\def\ee{\end{equation}}
\def\bea{\begin{eqnarray}}
\def\eea{\end{eqnarray}}

\begin{document}
\vspace*{4cm}
\title{PROTON LIFETIME FROM SU(5) IN EXTRADIMENSIONS}

\author{M.LAURA ALCIATI}

\address{ Dipartimento di Fisica `G.~Galilei', Universit\`a di Padova\\
INFN, Sezione di Padova, Via Marzolo~8, I-35131 Padua, Italy}

\maketitle\abstracts{We perform a detailed analysis in order to estimate the proton lifetime in the context of simple supersymmetric SU(5) model with an extradimension compactified on the orbifold $S^1/(Z_2\times Z_2')$. }

%\section{Guidelines}
%\subsection{Producing the Hard Copy}\label{subsec:prod}
%The hard copy may be printed using the procedure given below.
%You should use
%two files: \footnote{You can get these files from
%our WWW pages at {\sf http://www-lpnhep.in2p3.fr/moriond97/VHE$\_$Phenomena/}.}
%\noindent {\em moriond.sty} --- the style file that provides the higher
%level latex commands for the proceedings. Don't change these parameters.
%\noindent {\em moriond.tex}

%%%%%%%%%%%%%%%%%%%%%%%%%
%                       %
%       my paper        %                            
%                       %
%%%%%%%%%%%%%%%%%%%%%%%%%

The introduction of extradimensions can solve in a very elegant way some problems arising in the context of simple supersymmetric SU(5) model. We know that the latter is actually ruled out (or at least tightly constrained) because of the presence of dangerous dimension 5 operators that cause too fast proton decay, in contrast with present stringent limits on the proton lifetime set in many channels.
\\
\noindent From the theoretical point of view a real difficulty of minimal GUT model is the so called doublet-triplet splitting problem, that is how to explain light electroweak Higgs doublets but heavy Higgs triplets, both occurring in the same gauge multiplet. 

With the introduction of an extradimension compactified on an orbifold, we can solve these two problems in a simple and elegant 
way , since we can create a {\it KK}-tower of particles, whose zero modes are only the Standard Model gauge bosons and the Higgs doublets, while the SU(5) but not SM gauge bosons and the Higgs triplets get heavy masses of order of the compactification scale $M_c\equiv 1/R$. 
So the orbifold, not only solve the doublet-triplet problem, without the introduction of an {\it ad hoc} scalar potential, but also mediate the gauge symmetry breaking, because zero modes do not form a full multiplet of the gauge group. 

\noindent The automatic doublet-triplet splitting is also accompanied by the disappearance of dimension five operators for p-decay, after the introduction of an $U(1)_R$ symmetry.
Therefore proton decay can only proceed through dimension six
operators, that is through the interactions between fermions and gauge bosons.

The main ingredients in order to estimate proton lifetime are the compactification scale, as far as the decay rate scales as the fourth-inverse power of it, and the fermionic sector.

\noindent We evaluate $M_c$ from a next-to-leading order analysis of the gauge coupling unification, including two-loop running, heavy thresholds coming from KK particles, light threshold from Susy particles and $SU(5)$ violating terms, due to the presence of kinetic terms at the SM brane, allowed in principle by the theory.
We also account for experimental error, dominated by $\alpha_3$, but the largest error bar is the linear sum of this one and the theoretical one from the unknown $SU(5)$ violating terms. The probability is almost equal in this interval, because we have no reason to choice a value, rather than another. This big uncertainty make the compactification scale to lie in a relatively large interval: 

\begin{center} $M_c\approx 10^{14}\div 10^{16}$ GeV
\end{center}

As far as concerns the fermionic sector, we analyse 3 different options, according to different localization of matter fields on the brane or in the bulk.
OPTION $0$ is the simplest $SU(5)$ five-dimensional realization, since all matter is localized on the $SU(5)$ brane, so that the couplings between gauge bosons and fermions are essentially the same of conventional model. This option is almost ruled out by p-decay.

\noindent In the other two options the observed hierarchies among fermion masses arise from different localizations of matter fields. Only brane fields can contribute to p-decay, and this bring to huge flavor suppression. The dominant decay mode is $p \rightarrow K^+ \nu$, since this final state has the smallest flavor suppression.

\noindent We notice that there is a huge theoretical uncertainty that spreads over many order of magnitudes, due to the compactification scale $M_c$, which is only know up to about two order of magnitudes.
Since the proton lifetime scales as $M_c^4$,
this corresponds to an uncertainty of more than eight order of magnitudes
on the inverse rates.

%%%%%%%%%%%%%%%%%%%%%%%%%%%    grafici   %%%%%%%%%%%%%%%%%%%%%%%%%%%

\begin{figure}
\centerline{
\epsfig{figure=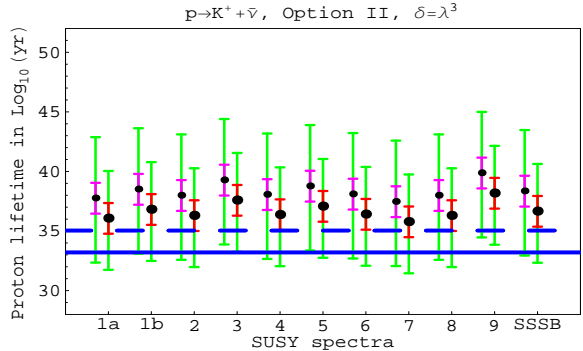,height=5cm,width=9cm}}
%\includegraphics[width=12.0cm]{plotb3.eps}
%\end{center}
\caption{Inverse decay rate of $p \rightarrow K^+ {\bar \nu}$ (summed over all neutrino channels) versus SUSY spectrum, in Option II ($5_{1,2,3}$ and $10_3$ are brane fields, while $10_1$ and $10_2$ live in the bulk), 
for $\delta=\lambda^3$( $\lambda$ is the cabibbo angle). 
The solid line is the current experimental lower bound,
and the dashed line is the aimed-for future sensitivity}
\label{fig5}
\end{figure}
%%%%%%%%%%%%%%%%%%%%%%
%                    %
%   Acknowledgments  %
%                    %
%%%%%%%%%%%%%%%%%%%%%%
\section*{Acknowledgments}
I'd like to thank Ferruccio Feruglio for the essential and kind collaboration. I thank also Lin Yin and Alvise Varagnolo for their important support. A special thank also to the organizers of 'Recontres de Moriond', for the beautiful atmosphere of the conference.  This project is partially
supported by the European Program MRTN-CT-2004-503369.
%%%%%%%%%%%%%%%%%%%%%%
%                    %
%     References     %
%                    %
%%%%%%%%%%%%%%%%%%%%%%


\section*{References}
\begin{thebibliography}{99}
\bibitem{LFLA}
M.~L.~Alciati, F.~Feruglio, Y.~Lin and A.~Varagnolo,
  %``Proton lifetime from SU(5) unification in extra dimensions,''
  JHEP {\bf 0503} (2005) 054
  [arXiv:hep-ph/0501086] and reference therein

  %%CITATION = HEP-PH 0501086;%%



\end{thebibliography}
\end{document}